Paper number ITS-2736

# Real scenario and simulations on GLOSA traffic light system for reduced CO2 emissions, waiting time and travel time

Marie-Ange Lebre[1,2*], Frédéric Le Mouël[2], Eric Ménard[1], Alexandre Garnault[1], Benazouz Bradaï[1], Vanessa Picron[1]

1. VALEO Advanced Technology Development, F-94000 Creteil, France
2. Université de Lyon, INSA-Lyon, CITI-INRIA Lab, F-69621 Villeurbanne, France

marie-ange.lebre,frederic.le-mouel@insa-lyon.fr // eric.menard,alexandre.garnault@valeo.com

**Abstract**

Cooperative ITS is enabling vehicles to communicate with the infrastructure to provide improvements in traffic control. A promising approach consists in anticipating the road profile and the upcoming dynamic events like traffic lights. This topic has been addressed in the French public project Co-Drive through functions developed by Valeo named Green Light Optimal Speed Advisor (GLOSA). The system advises the optimal speed to pass the next traffic light without stopping. This paper presents results of its performance in different scenarios through simulations and real driving measurements. A scaling is done in an urban area, with different penetration rates in vehicle and infrastructure equipment for vehicular communication. Our simulation results indicate that GLOSA can reduce CO2 emissions, waiting time and travel time, both in experimental conditions and in real traffic conditions.

## 1. Introduction

Advances in vehicular communications have led to the advent of cooperative intelligent transportation systems (ITS). These systems use wireless communication defined by the IEEE 802.11p standard. This new field of communications led car manufacturers and research institutes to deploy applications and services improving traffic efficiency, road safety, reducing CO2 emissions, etc. Vehicular communication, also called V2X with V2V for vehicle to vehicle and V2I for vehicle to infrastructure is a growing topic of interest in the scope of Advance Driving Assistance Systems (ADAS) and Automated Vehicles. It increases the field of view of the vehicles about their environment, enabling them to know a situation in advance, for example traffic jam, black ice, or other vehicles presence and intentions.

Vehicular communication also enables cooperation between agents, such as platooning for cruising vehicles, or green light requests for emergency vehicles approaching traffic lights. The targeted goal with the use of V2X in combination with ADAS is to increase safety and traffic efficiency. Currently applications on traffic control and reducing emissions are among the most promising according transportation authorities and car manufacturers. In this context of intelligent transportation, intersections equipped with traffic lights in urban areas are widely studied. Indeed, they represent significant and central nodes in an urban network in term of traffic control. The French funded public project Co-Drive aimed to validate a pre-industrialization approach towards a cooperative driving system between User, Vehicle and Infrastructure in order to suggest an intelligent, secure and calm route, for sustainable mobility [8]. The project duration was three years and ended on February 2014. The results of the project include prototype systems that detect incidents on the road and measure traffic data with the embedded devices, as a mobile traffic sensor.

These prototypes allowed sending data to traffic information service system and highway management system. With the data received, these systems were able to disseminate information about traffic or incidents to the on - coming vehicles before they reach it. The data transmission from vehicle to infrastructure used both VANET (Vehicular Ad Hoc Network) and cellular network. The second system developed in the framework of the Co-Drive project making the link between the vehicle and infrastructure is a cooperative traffic light system. It enables a safe crossing of the intersection with reduced $CO_2$ emission thanks to a Green Light Optimal Speed Advisory system (GLOSA). GLOSA has to inform the driver of the speed that he should maintain to pass the next traffic light with a green phase. Technically, cooperative traffic lights send information to vehicles thanks to V2X communications, so that vehicles can process it, calculate the optimal speed, and advise the driver. GLOSA is included in ETSI TC ITS Basic Set of applications, which lists V2X applications that can be deployed after the standardization completion.

Our investigation targeted on producing results using GLOSA on Valeo demonstration car in a real scenario. Positive results on $CO_2$ emission reduction and stop time reduction have been obtained. We verify these results with a similar scenario through the Simulator of Urban Mobility (SUMO). Then we scale the situation in two simulated urban areas. The first urban area allows seeing the performance of GLOSA according to the rate of penetration of vehicles and infrastructure, and the second simulated scenario allows evaluating the performance of GLOSA in a real situation. Indeed, we consider the existing travels of shuttles available for Valeo employees in Bobigny, bringing them from the train station to the Valeo site. This last test allows showing the advantages of this application in a daily situation with a reduction of travel time, waiting time, $CO_2$ emissions and also fuel consumption. In fact, the study of fuel consumption is not really necessary because the $CO_2$ emissions of a car are directly proportional to the quantity of fuel consumed by an engine.

The paper is structured as follows: at first, related work is presented on applications for the management of intersections in an ITS context. Then, our GLOSA system is presented in details. Finally, different simulation runs and a test in real condition are presented. Different penetration rates in equipment will be applied. We will show, through different scales of demonstrations, the power of this kind of applications in the field of intelligent transportation. Our results indicate that our GLOSA system could help reducing $CO_2$ emissions, waiting time and travel time according to the GLOSA equipped light penetration.

## 2. Related Work

Among all available solutions in urban environments to control intersections, the reservation-based centralized control system has been found to work best for urban intersections with high traffic demand thanks to its mechanism of maximizing the intersection capacity. Currently, this process is widely used in cities. Wu et al. [14] develop their own tool rather than using standard traffic simulation packages such as VISSIM, AIMSUN, or PARAMICS because the standard packages do not allow vehicles to be controlled individually. However, this process is very costly in term of equipment and maintenance.

In research fields, different protocols to control intersections exist. We can classify them into five groups: (1) Reservation (R): the vehicle sends a reservation request of space and time at the intersection, the server accepts or not this one; (2) Cooperative Adaptive Cruise Control (CACC): at intersections the server detects the conflicts and sends acceleration or deceleration requests to be respected by vehicles to avoid collisions; (3) Sequence (S): the intersection is controlled by sequencing the accesses of vehicles; (4) GLOSA: the vehicle adapts its speed according to data received from the intersections; (5) Traffic Light Control (TLC): the different phases of the traffic are controlled and vary according to the real traffic flow. In the first four groups, interactions exist between vehicles and the intersections thanks to sensors or wireless communications. In the last one, the control is done only by the intersection.

Hausknecht et al. [10] (R) use an autonomous intersection management approach. The manager predicts potential collisions by dividing the intersection into a grid and reserving space and time for each autonomous vehicle. As vehicles are near an intersection, they request their desired movement and the intersection manager determines when drivers can safely make the turn and sends back a reply.

Zhang et al. [16] (CACC) propose a fuzzy system, including generic algorithms. It allows processing uncertain information to be simply represented as simple rules. For example, traffic states such as speed, occupancy are grouped into simple categories such as high, medium, and low. For example, a rule can be: if the occupancy is high, speed is low, then congestion is positive. The originality of their article is that both autonomous and manually driven vehicles are considered, so various aspects of human behavior influence their system but they consider that intersections have no traffic light.

Ahamane et al. [2] (S) propose a right of way displayed on the board of non-autonomous vehicle approaching an intersection. They use Petri Net to model and compute an order for vehicles at an isolated intersection. Vehicles can communicate with each other wirelessly.

Youcef et al. [15] (TLC) present an adaptive traffic light control system based on Wireless Sensor Network (WSN) for controlling the traffic flow sequences on single and multiple intersections. The system consists of WSN -or TSNs (traffic sensor nodes) - installed on roadside that periodically collects the traffic data (vehicle speed and length), and sends it to control box (base station) that run control algorithms. Faye et al. (TLC) [7] introduce an algorithm named TAPIOCA, this distributed algorithm determines the green light duration and sequence for one isolated intersection or a set of adjacent intersections. They use wireless sensors to aggregate information and send them to the intersection manager, which uses this algorithm. Their solution is locally centralized.

Many other articles use GLOSA, to present its benefits through simulation. The present article aims at completing the article by Bradaï et al. [4] which present the gains of a GLOSA system on a real vehicle, by making the link between the measurements on the vehicle and the simulations to scale them up. The first implementation of GLOSA presented by Katsaros et al. [11] highlights its performance using an integrated cooperative ITS simulation platform in a simple urban scenario. Gajananan et al. [9] perform large-scale GLOSA system by creating a middleware framework, which integrates a traffic and communication network simulators with a multi-user driving simulation and focusing on the human response. Krajzewicz et al. [12] present the article on GLOSA used in the European project DRIVE C2X [1].

Cooperative intersections and smart intersection (without traffic light) systems are promising in terms of traffic efficiency and $CO_2$ emission reduction but maybe too far from the current standardization process and an actual deployment. In the case of such deployment with this powerful systems, that are able to manage completely the handling of intersections by vehicles, the question of the responsibility in the event of accident due to malfunction of a part of the system remain unclear, as it is a cooperative system between several vehicles and the infrastructure. Some projects also foresee the possibility of a deployment of wireless sensor network for intersection, but this is costly compared to a deployment of wireless communication systems in terms of tools and standards. Finally, the possibility of a centralized system brings several problems in terms of safety if a single failure occurs. Moreover, the centralized solutions have difficulties to treat information in real time due to the important quantity of data in a vehicular network. These issues have to be studied for the best understanding of the system and its implications in case of a massive deployment. The GLOSA systems seem to be an interesting first step in ADAS systems based on V2X communication with infrastructure, before the use of information by autonomous vehicles to handle intersections.

## 3. GLOSA: Simulation and Experimental Results

*3.1 GLOSA*

The objective of GLOSA systems is to provide to the driver the optimal speed to cross the next intersection with a green phase. The intersection needs to be equipped with a communication device and be interfaced with the intersection controller. This way, the intersection sends data to the approaching vehicles such as the intersection topology, the location, the traffic lights current phases and their duration. The on-board system processes these data to calculate the optimal approaching speed of the vehicle to reach the intersection with a green light. GLOSA systems improve traffic efficiency by: (1) reducing stop times, (2) bettering the fluidity of the traffic, (3) giving anticipating data which improve the safety, (4) reducing $CO_2$ emissions, fuel consumption and reducing waiting time and travel time.

This kind of system also helps the vehicles reaching the Green Wave if several traffic lights are coordinated as Faye et al. have done with wireless sensor networks [6]. It ensures a continuous flow of vehicles. Moreover, this system can be useful for emergency vehicles to request a right of way if traffic lights are able to communicate.

*3.2 Real and simulated scenario with one vehicle*

In this section, we present a simple scenario with one vehicle on a ring track. The test and measurements have been performed with an actual Valeo demonstration vehicle. The same scenario has been implemented with a simulation tool to compare the theoretical and practical results, and to highlight the reliability of the simulation model.

**Equipment for the real scenario.** A first implementation of a GLOSA algorithm was done based on the algorithm presented by Katsaros et al. [11]. During test drives, users gave good feedback about the feature, but it appeared that giving the driver textual information for GLOSA was not an intuitive way of advising about speed. A second implementation based on this algorithm targeted a better Human Machine Interface through quick reading and comprehension of the information with intuitive driving function, giving the driver the speed ranges corresponding to the phases of the next traffic light. This HMI is shown in the Figure 1 (b). The traffic light has communication devices and can transmit: identity, timestamp, latitude, longitude, current phase, remaining time of the current phase, green/amber/red phase duration. The traffic light control is embedded in a Raspberry Pi computer and sends the above data to the V2X equipment through Ethernet connection. The V2X equipment is composed of an IEEE 802.11p compliant WiFi router and an antenna mounted on the traffic light, to transfer the data to the approaching vehicles.

The Vehicle is a Peugeot 207 [8] equipped with a V2X communication device to receive data from traffic lights. A Smartphone is also used as the sensing unit that receives data from the traffic light via the V2X device embedded in the vehicle. It can retrieve localization and speed data with the embedded GPS system. The Human Machine Interface finally delivers the Green Light Optimal Speed to the driver.

The experiment took place on a speed ring test track, with 2 set-up traffic light as shown in the Figure 1 (c). The two traffic lights had the same cycle: Red 30s / Amber 2s / Green 25s The two traffic lights were not synchronized; they were 1500 meters far from each other. No other vehicle is involved in the test. Now we present the simple scenario with a simulator.

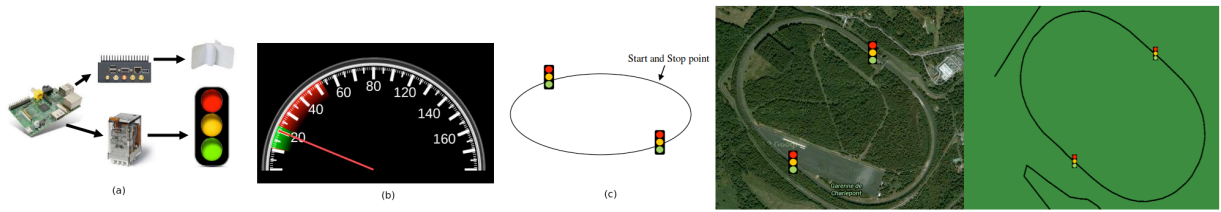

**Figure 1 – (a) Traffic light equipment scheme (b) GLOSA HMI (c) Traffic lights set-up on the test track and the simulated test.**

**Environment and hypothesis for the simulated scenario.** The simulator used is the Simulator of Urban Mobility – SUMO version 0.22 ([3]). It is an open source microscopic and continuous road traffic simulation package. The Traffic Control Interface (TraCI) [13] enables the control of vehicles while the simulation is running. With this access to a running road traffic simulation, it allows to retrieve characteristics of simulated objects and to manipulate their behavior "on-line". We use it to implement GLOSA in our simulation. The infrastructure parameters can be adjusted the same way, what can be seen in the Figure 1 (c).

*3.2.1 Results*

The measures presented show the effects of GLOSA for crossing an intersection with traffic lights in terms of speed and $CO_2$ emission. Figure 2 (a) and (b) displays the speed profile with and without GLOSA in the real (a) and simulated scenario (b) for the same situation where the vehicle approaches the traffic light. The two figures show that there is no stop time with the GLOSA system in the presented case. The speed profile of the simulated situation is less smooth than in the reality. Indeed, the vehicle directly takes the recommended speed by GLOSA while the driver will gradually decrease his speed. The results obtained and the figures presented show that both situations are very similar. Figures 2 (c) and (d) show the difference in $CO_2$ emission for the same situation for both scenarios: real (c) and simulated (d). The $CO_2$ emission reduction is due to the facts that there is no unnecessary acceleration and possibly avoidance of stop times when using the GLOSA system, and that the acceleration phase after the traffic light is reduced.

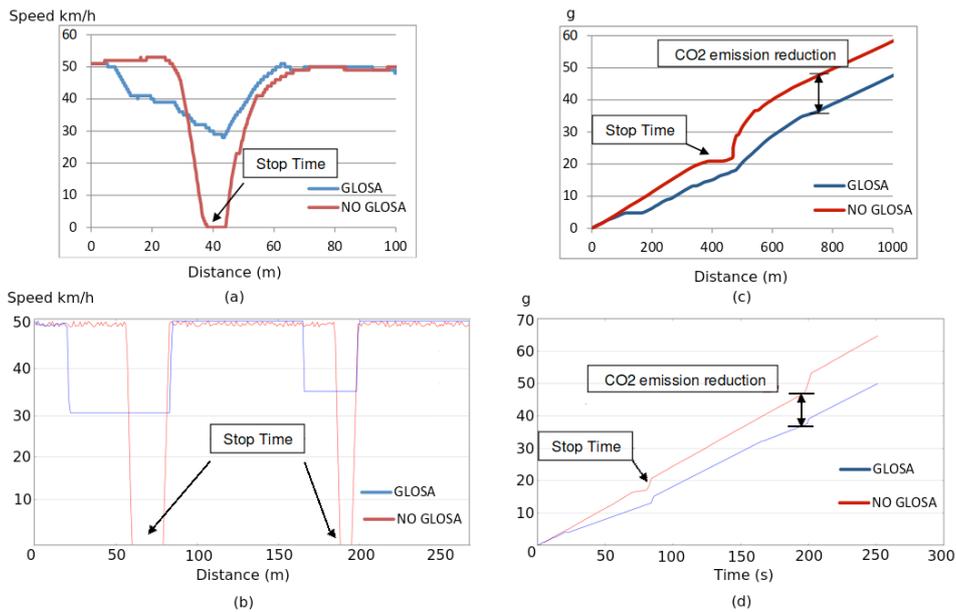

**Figure 2 - Speed (km/h) versus time (s): real scenario (a) and simulated scenario (b) // Cumulated CO2 emission (g) versus distance travelled (m) in the real scenario (c) and versus time (s) in simulated scenario (d)**

Simulated and real scenarios have the same profile. The difference with and without GLOSA is significant and similar in the two cases. There is a reduction of 11% in the simulated scenarios and 13% in the real case. The difference obtained can be the result of the difference in acceleration and acceleration profiles, and that the simulation may not take into account all the vehicle parameters regarding its powertrain system. This difference in the results remains minor regarding the scenario and the global results obtained.

The comparison between the simulated scenario and the real one, confirms that the simulations implemented are reliable regarding this system.

*3.3 Simulated scenario with traffic*

Now we consider the same situation as the previous one but this time with other vehicles on the road. Regarding the infrastructure, three situations are simulated: traffic lights are not equipped, then one traffic light is equipped, and, finally, the two traffic lights are equipped. The first situation corresponds to the situation without GLOSA. Then we consider different percentages (%) of equipped vehicles: [10, 20, 30, 40, 50, 60, 70, 80, 90, 100]. In this simulation: 100 vehicles enter the track every 10 seconds.

For each vehicle, we compare the mean of CO2 emissions and waiting time with the means without GLOSA, giving the gains to expect with GLOSA system in this situation. We obtain the 3D graph in Figures 3, which shows the benefit of GLOSA. We observe a reduction of 10% in CO2 emissions with a penetration rate of 100% for vehicles and infrastructure. We also notice that, with 40% of equipped vehicles and 50% of equipped traffic lights, there is a reduction of 5% in CO2 emissions and 30% of waiting time. It means that as soon as we equip half vehicles and traffic lights there is a notable gain.

We observe that vehicles never stop with 100% equipped vehicles and infrastructure; there is no waiting time in the simulation on the track. Indeed, there is no traffic jam in this scenario, consequently each vehicle, equipped or not, can benefit from the GLOSA system.

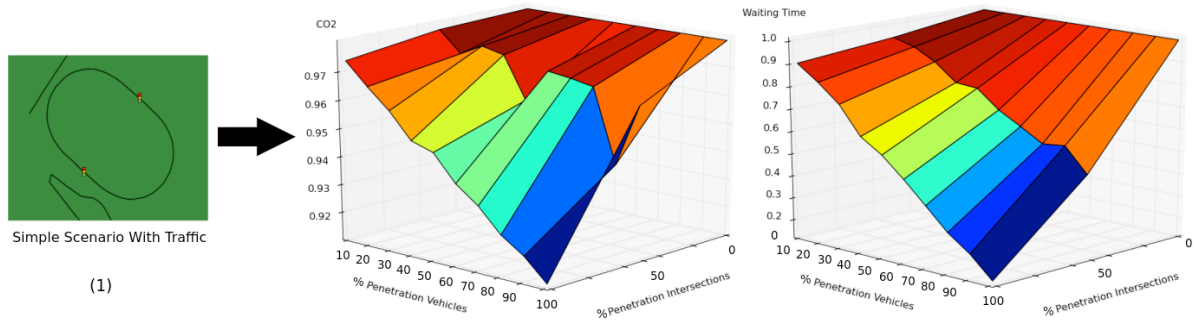

**Figure 3 - Results for the simple scenario with traffic**

*3.4 Scaling scenarios*

Now, we present two simulations in an urban area: the first one is a typical case study – the Manhattan map and the second one is a real map in Bobigny (France) – the "Valeo Shuttle".

*3.4.1 Manhattan map*

The map of Manhattan is a 5x5 grid with nine traffic lights in the center. The length of the lane between two traffic lights is 900 meters. At each step, 200 vehicles appear at random locations on the road network and travel to a random destination with the shortest path, computed with Dijkstra's algorithm. We apply different percentages for wireless equipment for traffic lights (0% to 100%), and we do the same for the 200 vehicles (10% to 100%).

At the end of the simulations, we compare for each vehicle the mean of CO2 emissions, waiting time and travel time with the means without GLOSA (see the ratio on Y-axis in the 3D graphs in Figure 4). As in the previous simple scenario, the traffic jam is not significant. We find that with a real network (next section) the reduction of waiting time is worse than in a simple grid where vehicles do not suffer in the speed variation. In the results, we observe, with a penetration rate of 100% for vehicles and infrastructure, a reduction of 10.5% in CO2 emissions, a gain of 80% in waiting time and 2.5% in time travel. The maximum gain of 3% in travel time is obtained with a rate of 50% for the traffic lights and 100% for the vehicles. It means that a total deployment is not necessary for this configuration. As the previous simulations (the simple scenario with traffic), with 40% of equipped traffic lights and 50% of equipped vehicles we observe 6% of reduction of CO2 emissions and 30% of reduction of waiting time. Therefore, results are relatively stable even if the urban configuration is different and a partial deployment of the wireless equipment shows the great potential of GLOSA.

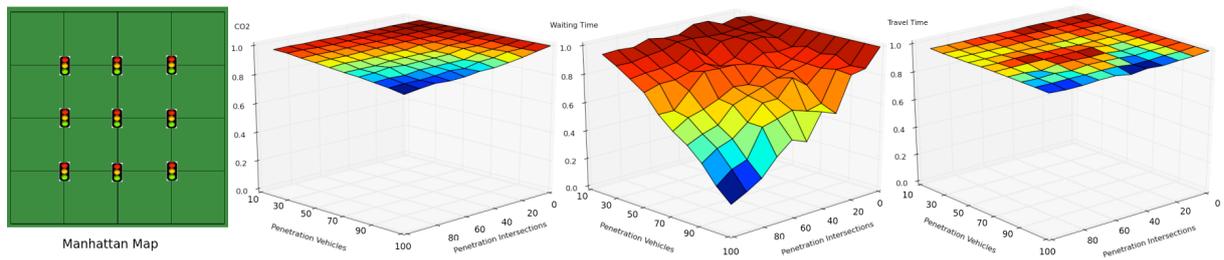

Figure 4 - Results for Manhattan map scenario

*3.4.2 Valeo shuttles*

This last scenario can be considered as a real life scenario, we focus on shuttles transporting Valeo employees from the train station to Valeo Bobigny site (see the maps in Figure 5). The maps measure 2.7 km x 3 km, the red line represents the distance travelled by the shuttles in both directions. We add different vehicles densities to see the effects of a GLOSA system in a real life situation. As previous simulations, various penetration rates are applied to the vehicles and shuttles (10 to 100 %) and the traffic lights (0 to 100 %).

In each simulation, a departure of a shuttle is in each direction (Valeo site <-> Train station) with different frequency during 600 seconds. For instance in the first simulation with low density, two shuttles enter the network (one at the Valeo site and the other at the train station) every 200 seconds during 600 seconds, therefore there are 6 shuttles. We add vehicles on the same axis as the shuttles during 600 seconds. Finally, we add vehicles with random travels in the network (see the table 2 for all the details). There are 154 vehicles in the first simulation (Low), 309 vehicles in the second (Medium) and 512 vehicles in the third (High).

| Location | Type | Duration(s) | Frequency(s) | | | Number | | |
|---|---|---|---|---|---|---|---|---|
| | | | L | M | H | L | M | H |
| Valeo site <-> Train station | Shuttles | 600 | 200 | 100 | 50 | 6 | 12 | 24 |
| Valeo site <-> Train station | Vehicles | 600 | 60 | 30 | 15 | 20 | 40 | 80 |
| At random on the map | Vehicles | | 1 | 1 | 1 | 128 | 257 | 408 |
| | Total | | | | | 154 | 309 | 512 |

Table 1 - Parameters of the simulation with different vehicles densities, L for Low density, M for Medium and H for High density.

Results are presented in Figure 5. A small reduction (below 1 %) for the travel time is observed once that the vehicles and intersections are equipped. With a low vehicles density on the network, the travel time is improved by 1.95 %, with 100 % of the vehicles and 60 % of the intersections equipped. So the network needs an important number of equipped vehicles, in case of low-density, to have significant travel time improvement.

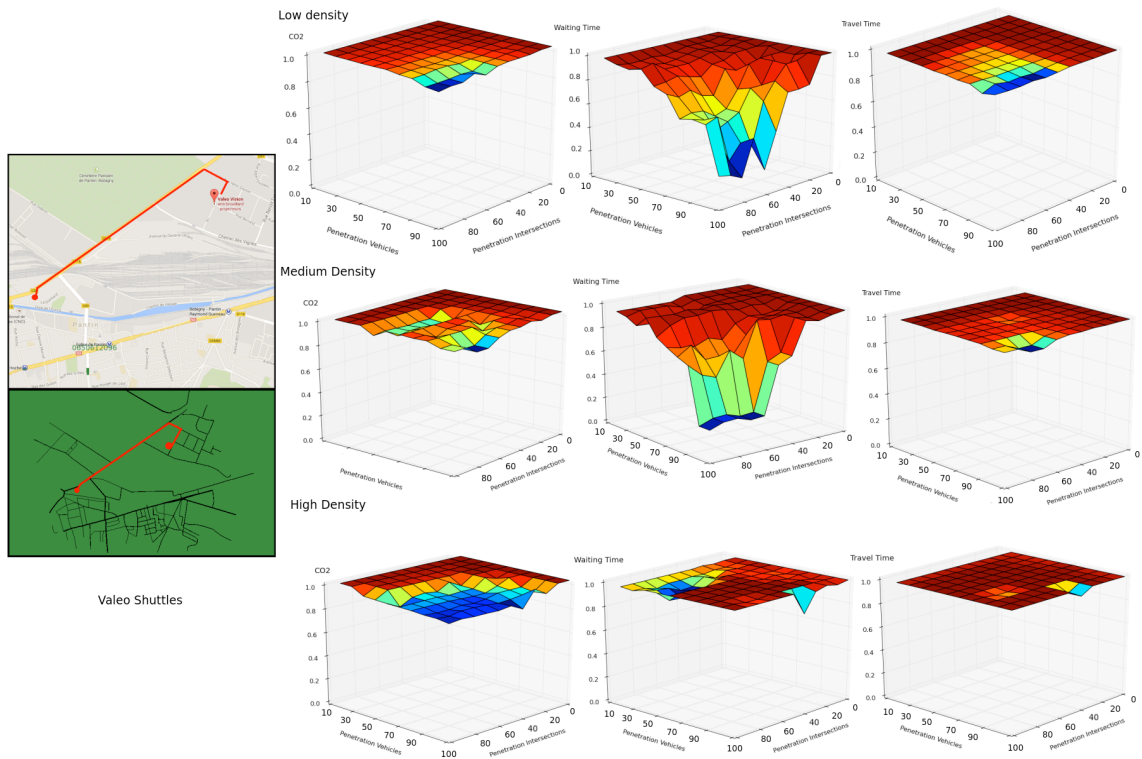

**Figure 5 - Results for scenario with Valeo Shuttles**

In fact when the circulation is fluid, saving time is not obvious with low equipment. By contrast in the situation with a medium density, the reduction of the travel time is around 2.05 % with only 50 % of equipped vehicles and 50% for the intersections. In the case of high density, there is an improvement of 5.6% with 90% of the vehicles equipped and 40% of the intersections equipped. There is also a reduction of 1.3% with 70% of equipped vehicles and intersections. Overall, no improvement of travel time can be performed with a high density. Eckhoff et al. [5] show that, with denser traffic, the performance of GLOSA is deteriorated. We observe longer waiting and travel times and more frequent stops for all vehicles.

For the waiting time the reduction increases very quickly in function of the equipment rate with 16.5 % of reduction with 50 % of vehicles and 60 % of intersections for a low density and a reduction of 32 % with 70 % of vehicles and 60% of intersections equipped for a medium density. The waiting time has a small reduction (below 2%) in the case of high density for all percentage of equipment. Except for two set of values: with only 40 % of the vehicles equipped and 80 % of the intersections, there is a reduction of 8.9%, and with 100 % of the vehicles equipped and 40 % of the intersections, there is a reduction of 18.17%.

Finally, the $CO_2$ reduction is not positive or even slightly negative for low equipment in case of low density. The gain becomes positive only for 90 % of vehicles and 90 % of intersections: 1.88 %. In the case of a medium density, a reduction was observed once that 50 % of the vehicles and 40 % of the intersections are equipped. On the contrary, the reduction increases with the equipment in the case of high density with a reduction of 4.5 % and 10 % with 70 % and 100% of equipped vehicles respectively and 50% and 40% of equipped intersections respectively.

In the table below we present the equipment rate where reductions in travel time in waiting time and CO2 emissions are best for different vehicles densities. We notice that the maximum reduction in the case of high density is effective with only 40 % of equipped intersections.

|  | Vehicles equipment (%) | | | Intersections equipment (%) | | | Maximum Reduction (%) | | |
| --- | --- | --- | --- | --- | --- | --- | --- | --- | --- |
|  | L | M | H | L | M | H | L | M | H |
| Travel Time | 90 | 90 | 90 | 100 | 70 | 40 | 3.89 | 6.21 | 5.66 |
| Waiting Time | 100 | 90 | 100 | 80 | 100 | 40 | 82.4 | 75.63 | 18.17 |
| CO2 | 100 | 90 | 100 | 100 | 70 | 40 | 2.81 | 7.37 | 9.91 |

**Table 2 - Rate of equipment for a maximum rate of reduction for each parameter.**

**4. Conclusion**

In this paper, a GLOSA system has been presented and tested with simulations and real scenarios. The comparison between a real scenario and the simulated scenario shows the reliability of the simulations and allows a scaling in an urban area. GLOSA can help reducing CO2 emissions, waiting time and travel time both in experimental conditions and in real traffic conditions. Performances do not depend on the physical urban area topology. Moreover, GLOSA has an impact on non-equipped vehicles, that must undergo the behavior of equipped vehicles slowing down when approaching an intersection, but that also benefit from it at the same time. Once the infrastructure and the vehicles begin to be equipped to 50%, the results are significant in each urban scenario. It would be interesting to evaluate the gain of cooperative traffic lights and the impact on user's experience and acceptance, and also on experience and acceptance of non-equipped vehicles drivers. Finally, the impact of the traffic jam on the system GLOSA has to be studied in details.